\documentclass[acmsmall,screen]{acmart}

\setcopyright{rightsretained}
\acmDOI{10.1145/3808190}
\acmYear{2026}
\acmJournal{PACMSE}
\acmVolume{3}
\acmNumber{FSE}
\acmArticle{FSE183}
\acmMonth{7}
\acmSubmissionID{fse26mainb-p1886-p}
\received{2026-02-25}
\received[accepted]{2026-03-24}

\AtBeginDocument{
  \providecommand\BibTeX{{
    \normalfont B\kern-0.5em{\scshape i\kern-0.25em b}\kern-0.8em\TeX}}}
    
\usepackage[T1]{fontenc}
\usepackage{textgreek}
\usepackage{fancyvrb}
\usepackage{multirow}
\usepackage{colortbl}
\usepackage{longtable}
\usepackage{booktabs}
\usepackage{ifthen}
\usepackage{bbding}
\usepackage{makecell}
\usepackage{threeparttable}
\usepackage{amsmath}
\usepackage{stfloats}
\usepackage{xspace}
\usepackage{paralist}
\usepackage{enumerate}
\usepackage[linesnumbered,ruled,vlined]{algorithm2e}
\usepackage{array}
\usepackage{float}
\usepackage{tikz}
\usepackage{calc}
\usepackage{listings,amsfonts}
\usepackage{pifont}
\usepackage{url}
\usepackage{epsfig, endnotes}
\usepackage[normalem]{ulem}
\usepackage[misc]{ifsym}
\usepackage{subcaption}
\usepackage{cleveref}
\usepackage{tcolorbox}
\usepackage{python}
\usepackage{enumitem}
\usepackage{adjustbox}

\lstdefinelanguage{json}{
    morestring=[b]",
    morecomment=[l]{//},
    morecomment=[s]{/*}{*/},
    stringstyle=\color{red},
    commentstyle=\color{gray},
    keywordstyle=\color{blue}\bfseries,
    morekeywords={true,false,null},
    literate={\ }{{\ }}1 
}

\definecolor{codegray}{rgb}{0.5,0.5,0.5}
\definecolor{codeblue}{rgb}{0.0,0.0,0.6}
\definecolor{codegreen}{rgb}{0.1,0.5,0.1}
\definecolor{codepurple}{rgb}{0.58,0,0.82}
\definecolor{backcolour}{rgb}{0.97,0.97,0.97}

\definecolor{customblue}{HTML}{006ca6}
\definecolor{customgreen}{HTML}{009264}
\definecolor{custombrown}{HTML}{ff3d00}

\AtEndPreamble{

}
\newcommand{\find}[1]{
\begin{tcolorbox}[leftrule=0.5mm,toprule=0mm,bottomrule=0mm,left=0.7pt,right=0.7pt,top=0.2pt,bottom=0.2pt]
\em #1
\end{tcolorbox}
}

\title{Not All RAGs Are Created Equal: A Component-Wise Empirical Study for Software Engineering Tasks}

\author[Q. Ke]{Qiang Ke}
\orcid{0009-0003-2520-881X}
\affiliation{
  \department{Hubei Key Laboratory of Distributed System Security}
  \department{Hubei Engineering Research Center on Big Data Security}
  \department{School of Cyber Science and Engineering}
  \institution{Huazhong University of Science and Technology}
  \city{Wuhan}           
  \country{China}
}
\email{qiangke@hust.edu.cn}

\author[Y. Zhao]{Yanjie Zhao}
\orcid{0000-0001-8793-5367}
\authornote{Corresponding author (Yanjie\_Zhao@hust.edu.cn).} 
\affiliation{
  \department{Hubei Key Laboratory of Distributed System Security}
  \department{Hubei Engineering Research Center on Big Data Security}
  \department{School of Cyber Science and Engineering}
  \institution{Huazhong University of Science and Technology}
  \city{Wuhan}           
  \country{China}
}
\email{Yanjie_Zhao@hust.edu.cn}

\author[H. Leng]{Hongjin Leng}
\orcid{0009-0007-3607-8204}
\affiliation{
  \institution{Xiamen University Malaysia}
  \city{Sepang}
  \country{Malaysia}
}
\email{CYS2409015@xmu.edu.my}

\author[S. Zhao]{Shengming Zhao}
\orcid{0009-0007-5035-2206}
\affiliation{
  \institution{Fudan University}
  \city{Shanghai}
  \country{China}
}
\email{smzhao25@m.fudan.edu.cn}

\author[H. Wang]{Haoyu Wang}
\orcid{0000-0003-1100-8633}
\affiliation{
  \department{Hubei Key Laboratory of Distributed System Security}
  \department{Hubei Engineering Research Center on Big Data Security}
  \department{School of Cyber Science and Engineering}
  \institution{Huazhong University of Science and Technology}
  \city{Wuhan}           
  \country{China}
}
\email{haoyuwang@hust.edu.cn}

\begin{document}

\begin{abstract}

While Retrieval-Augmented Generation (RAG) is increasingly adopted to ground Large Language Models (LLMs) in software artifacts, the optimal configuration of its components remains an open question for software engineering (SE) tasks. The lack of systematic guidance forces practitioners into costly, ad-hoc experimentation. This paper presents a comprehensive, component-wise empirical study that dissects the RAG pipeline, evaluating over \textbf{21} distinct models and methods. Our study systematically isolates and evaluates \textbf{4} query processing techniques, \textbf{7} retrieval models spanning sparse, dense, and hybrid paradigms, \textbf{4} context refinement methods, and \textbf{6} distinct generators. We test these components on a suite of \textbf{3} core SE tasks: code generation, summarization, and repair. Our empirical findings reveal a crucial insight: the retriever-side components, particularly the choice of the retrieval algorithm, often exert a more significant influence on final system performance than the selection of the generator model. Strikingly, the classic lexical retriever \textit{BM25} demonstrates exceptionally robust performance across diverse tasks. Our analysis provides a practical, data-driven roadmap for researchers and practitioners, offering clear guidance on prioritizing optimization efforts when constructing effective RAG systems for software engineering contexts.

\end{abstract}



\maketitle

\section{Introduction}
\label{sec:introduction}

Retrieval-Augmented Generation (RAG) has become a widely adopted paradigm for enhancing Large Language Models (LLMs) in software engineering applications~\cite{lewis2021retrievalaugmentedgenerationknowledgeintensivenlp, gao2024retrievalaugmented}. By incorporating external knowledge bases such as project-specific documentation and code repositories, RAG reduces factual inaccuracies and improves the contextual relevance of generated outputs for code-related tasks~\cite{parvez2021retrieval, lu-etal-2022-reacc}.

However, constructing optimal RAG pipelines for software engineering tasks lacks systematic guidance. Software engineering presents unique challenges for RAG systems, including the structured nature of code \cite{alon2018code2ve}, diverse query types ranging from natural language questions to code snippets \cite{treude2011how, Sillito2012makes}, and the need to retrieve from heterogeneous knowledge sources such as documentation, API references, and code repositories \cite{zhou2023docprompting}. A RAG system consists of multiple interconnected components, each presenting numerous design options \cite{gao2024retrievalaugmented}. Practitioners must make critical decisions regarding query processing, retrieval algorithms, context refinement, and generator model selection. The absence of systematic, empirical evidence to inform these choices specifically for software engineering contexts results in a costly, iterative experimentation process.

The limited body of prior empirical work on RAG for code, while providing valuable initial insights, is constrained by a narrow scope. Existing studies have focused almost exclusively on the single task of code generation, investigating specific aspects of the retrieval stage, such as the effectiveness of different information sources like APIs \cite{gu2025what} or the quality and fusion of retrieved code snippets \cite{yang2025empirical}. Consequently, their findings, while important, are confined to a single component and a single task.

To address this limitation, this paper presents a systematic, component-wise empirical study of RAG pipelines for software engineering tasks. Our investigation methodically examines the pipeline to isolate and evaluate the impact of core architectural decisions. We aim to establish a data-driven understanding of how each component contributes to overall system performance. To guide our study, we formulate the following four research questions:

\textbf{RQ1: How does the query processing stage impact performance?} We investigate whether transforming the user's initial query, for instance, by simplifying or elaborating it, can enhance the retrieval of relevant information and improve the quality of the final output.

\textbf{RQ2: Which retrieval strategy is most effective for code tasks?} This question examines the core of the RAG pipeline by comparing different retrieval paradigms: sparse (e.g., keyword-based), dense (e.g., semantic-based), and hybrid approaches to determine which is best suited for the unique characteristics of code and natural language queries in software engineering.

\textbf{RQ3: What is the utility of the context refinement stage?} We evaluate post-retrieval techniques such as re-ranking and compression. This inquiry seeks to understand whether these optional steps effectively increase the signal-to-noise ratio for the generator or if they risk discarding critical information.

\textbf{RQ4: What is the relative importance of the generator versus the retriever?} The final question assesses the interplay between the retrieval and generation stages. We aim to determine whether final performance is more sensitive to the quality of the retrieved context or the intrinsic capabilities of the generative LLMs.

Through answering these questions, this work makes the following principal contributions:
\begin{itemize}
    \item We design and implement a \textbf{modular testbed} for RAG systems in software engineering (hereafter referred to as Code RAG) that decouples the pipeline's core stages, enabling flexible and reproducible component-wise experimentation.
    \item We conduct a \textbf{large-scale empirical study}, evaluating over \textbf{20} distinct models across \textbf{3} diverse code tasks on \textbf{4} datasets to build a foundational knowledge base for practitioners.
    \item We develop and open-source a \textbf{prototype adaptive RAG framework} that leverages our empirical findings to automatically recommend a near-optimal pipeline configuration based on task-specific features.
\end{itemize}

\section{Background and Motivations}
\label{sec:background}

The adoption of RAG has been pivotal in advancing the capabilities of LLMs for SE. By grounding generation in targeted, real-time context, RAG promises to make LLMs more reliable development partners. However, realizing this potential is hindered by several fundamental challenges that motivate our work.

\textbf{Challenge 1: Performance Instability and Context-Dependency.} A primary challenge is RAG's performance instability across different SE tasks. Our preliminary experiments, corroborated by findings in the literature~\cite{gao2024retrievalaugmented}, reveal that no single RAG configuration is universally optimal. The ideal choice of components is highly dependent on the task context. For instance, a pipeline tuned for code generation (\textit{text-to-code}) may fail on code repair (\textit{code-to-code}), meaning success on one benchmark rarely guarantees generalizability to diverse development scenarios.

\textbf{Challenge 2: The Expanding and Evolving RAG Design Space.} The RAG design space is exponentially growing. The classic pipeline forces choices among sparse, dense, or hybrid retrievers~\cite{Karpukhin2020DensePR, cormack2009reciprocalRF}, alongside post-retrieval refinements like re-ranking or compression~\cite{sun2024chatgptgoodsearchinvestigating, jiang2024longllmlinguaacceleratingenhancingllms}. Furthermore, while advanced, multi-step agentic paradigms are emerging~\cite{fan2024survey, arslan2024survey}, they fundamentally build upon these classic components. Their complexity makes a systematic, component-wise evaluation intractable without first establishing a foundational understanding of the core building blocks. Therefore, this study focuses on providing this essential empirical foundation for the classic RAG pipeline, which remains the most widely adopted and fundamental architecture in practice. The vast and poorly understood set of trade-offs within even this classic design makes it nearly impossible to select an optimal configuration without systematic guidance.

\textbf{Challenge 3: Lack of Systematic Empirical Guidance.} The majority of existing research introduces a novel component or a new architectural pattern and demonstrates its effectiveness on specific benchmarks. While such studies are valuable, they contribute to a fragmented understanding of the RAG ecosystem. Even comprehensive surveys that categorize RAG techniques~\cite{gao2024retrievalaugmented} often do not provide comparative empirical data on how different component \textit{combinations} perform across a \textit{variety} of SE task characteristics. Consequently, practitioners are forced to rely on costly trial-and-error. The field urgently needs an empirical roadmap mapping task features to optimal configurations.

\textbf{Our Approach.} To address these challenges, we undertake a \textbf{systematic, component-wise empirical study} mapping the RAG design space for SE. By rigorously evaluating and comparing a wide array of components in a controlled testbed, we aim to transform the ``alchemy'' of current RAG development into a more principled, scientific practice. Our goal is to provide the empirical foundations necessary for building more effective and reliable RAG systems in SE.

\section{Approach}
\label{sec:approach}

\begin{figure}[t]
    \centering
    \includegraphics[width=1\textwidth]{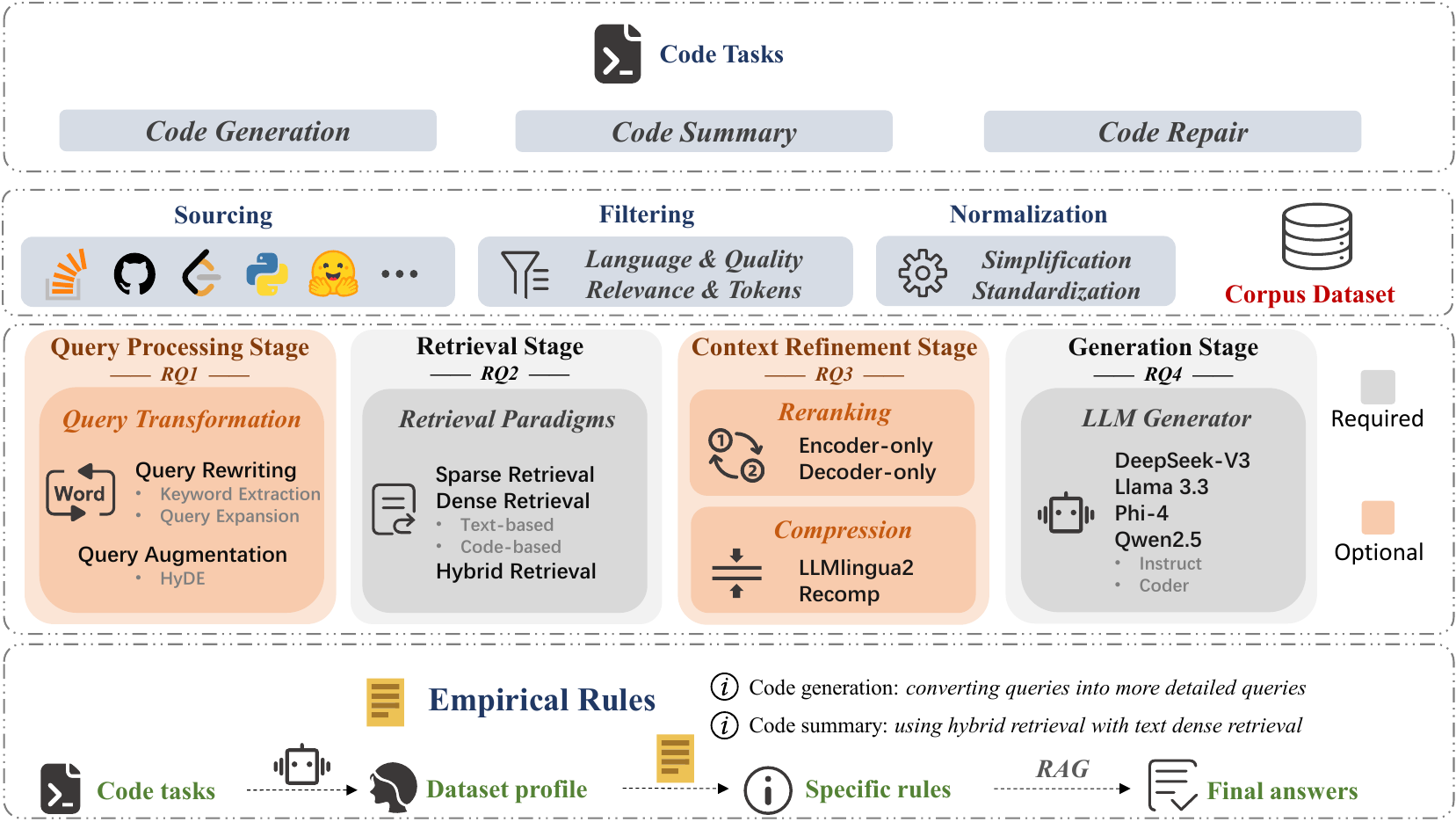}
    \Description{The overall architecture of our RAG framework.}
    \caption{The overall architecture of our RAG framework.}
    \label{fig:overview}
\end{figure}

This section details the methodology for our empirical study of RAG on code tasks. We first establish the foundations of our study by 
compiling a suite of code tasks and constructing a large-scale retrieval corpus. Next, we introduce our modular testbed, a plug-and-play architecture designed for rigorous, component-wise experimentation. Finally, guided by our experimental insights, we propose a prototype of an LLM-driven adaptive framework that recommends an optimal pipeline configuration tailored to specific task features. The overall architecture is illustrated in \autoref{fig:overview}.

\subsection{Foundations of the Empirical Study}

Our empirical study is built upon two foundations: a curated set of code tasks (processed in \autoref{sec:datasets}) and a large-scale retrieval corpus (constructed in \autoref{sec:corpora}). These were designed to systematically evaluate RAG components across diverse development scenarios.

\subsubsection{Code Tasks}

Our evaluation suite comprises three fundamental tasks. \textbf{Code Generation} is a \textit{text-to-code} task that synthesizes code from natural language, testing the retrieval of relevant algorithms and APIs. \textbf{Code Summarization}, a \textit{code-to-text} task, generates a concise description for a code snippet to assess semantic comprehension. Finally, \textbf{Code Repair} is a complex \textit{code-to-code} task where the system must understand buggy code and retrieve correct patterns to guide fixes. 

Crucially, rather than being isolated applications, these tasks were strategically selected as representative proxies for the three foundational input-output modalities in SE~\cite{chen2021evaluating, lu2021codexglue, Gazzola2017Automatic}. Consequently, empirical insights derived from this triad inherently generalize to a broader spectrum of SE scenarios, such as automated test generation (\textit{text-to-code}), documentation writing (\textit{code-to-text}), and code refactoring (\textit{code-to-code}).

\subsubsection{Retrieval Corpus}

A comprehensive, up-to-date corpus suitable for RAG across our diverse evaluation tasks is not readily available. We therefore constructed a new, large-scale knowledge source by integrating and processing data from five authoritative sources: Stack Overflow, GitHub, LeetCode, Python API docstrings, and Hugging Face. To ensure the integrity of our evaluation and prevent data leakage, we implemented a strict decontamination. We removed any document from our constructed corpus if it contained code that was an exact match to any code solution present in the test splits of our evaluation datasets. This process yielded a realistic and decontaminated corpus with a wide spectrum of coding styles and complexities, essential for a robust retrieval evaluation.

\subsection{The Modular Testbed for Code RAG}
\label{sec:stages}

At the heart of our empirical study is a modular testbed for Code RAG, designed with \textit{modularity} and \textit{plug-and-play} capability as its core principles \cite{wang2024searching}. This allows us to isolate, evaluate, and combine various RAG components seamlessly. The testbed decouples the RAG workflow into distinct, interchangeable stages, a fundamental architecture for conducting controlled experiments and attributing performance changes directly to the component being evaluated. The key stages are detailed below.

\subsubsection{Query Processing Stage}
\label{sec:query_processing}

This optional stage investigates whether refining raw queries enhances retrieval and downstream performance. Our testbed implements two primary refinement paradigms. To ensure robust findings unbiased by specific model characteristics, we employ the open-weight \textit{DeepSeek-V3} \cite{DeepSeekV2025DeepSeekVTR} and proprietary \textit{GPT-4o} \cite{openai2024gpt4o} as core engines. Their powerful reasoning capabilities ensure high-fidelity query transformations. The evaluation is detailed in RQ1 (\autoref{sec:rq1}).

\textbf{Query Transformation.} This approach rewrites the query to improve its semantic clarity and keyword relevance, inspired by recent work~\cite{Ma2023QueryRewriting}. We explore two distinct strategies:

\begin{itemize}
    \item \textbf{Simplified Query:} This method condenses the problem description into a concise format, retaining only core algorithmic keywords and essential constraints to reduce noise.
    \item \textbf{Elaborated Query:} In contrast, this method expands the query into a more structured format via chain-of-thought prompting~\cite{Wei2022ChainOfThought}. The model is instructed to include detailed I/O specifications, potential edge cases, and a pseudo-code sketch of a solution. This implicitly incorporates the core principle of \textit{HyDE}~\cite{Gao2022PreciseZeroShot} by proactively generating a \textit{hypothetical answer} within the query itself.
\end{itemize}

\textbf{Query Augmentation.} In contrast to rewriting, this method enriches the query by appending a known, correct solution. While the original \textit{HyDE} framework~\cite{Gao2022PreciseZeroShot} generates a hypothetical document, our implementation adapts this for an experimental setting by using the ground-truth answer. This serves as an oracle-based approach to create a semantically ideal query embedding, providing a practical upper bound on performance achievable through query enrichment.

\subsubsection{Retrieval Stage}
\label{sec:retrieval}
The Retrieval Stage is a mandatory step that recalls an initial set of candidate documents from the code corpus. Our framework systematically evaluates three distinct retrieval paradigms. For each paradigm, we select a suite of representative and commonly used models, which are detailed below and summarized in \autoref{tab:retrieval_models}. Their performance is analyzed in RQ2 (\autoref{sec:rq2}).

\textbf{Sparse Retrieval.} This paradigm relies on classic, non-deep-learning models that excel at exact keyword matching based on term frequency (TF) and inverse document frequency (IDF)~\cite{sparck1972statistical}. In our study, \textit{BM25}~\cite{Robertson1994simpleeffectiveapproximations} serves as a crucial lexical baseline, representing the robust performance of term-based search.

\textbf{Dense Retrieval.} This paradigm utilizes deep learning models for semantic search, encoding queries and documents into high-dimensional vectors (embeddings) within a shared space~\cite{Karpukhin2020DensePR}. We use the same model for both queries and documents to ensure a consistent mapping. Embeddings are indexed using the Faiss library~\cite{johnson2021billionscalesimilaritysearchgpus} with the \textit{IndexFlatL2} algorithm to ensure exact similarity search. We evaluate two categories of embedders:

\begin{itemize}
    \item \textbf{General-purpose Embedders:} Pre-trained on diverse text corpora~\cite{wang2024improvingtextembeddingslarge}, these models possess a broad understanding of natural language semantics.
    \item \textbf{Code-specialized Embedders:} Fine-tuned on large-scale source code~\cite{Feng2020CodeBERTAP}, these models are trained to understand the nuances of programming languages for better retrieval of functionally relevant code.
\end{itemize}

\textbf{Hybrid Retrieval.} This approach combines the outputs of sparse and dense retrieval to synthesize their respective strengths. We use the Reciprocal Rank Fusion (RRF) algorithm~\cite{cormack2009reciprocalRF}, which computes a new score for each document $d$ as follows:

\begin{equation}
\text{Score}_{RRF}(d) = \sum_{r \in R} \frac{1}{k + \text{rank}_r(d)}
\end{equation}

where $R$ is the set of ranked lists, $\text{rank}_r(d)$ is the rank of document $d$ in list $r$, and $k$ is a smoothing constant. In our experiments, we set $k=60$, following the established academic and industry standard~\cite{cormack2009reciprocalRF}, as it is mathematically proven to optimally mitigate the influence of lower-ranked outlier documents without requiring further parameter tuning. We fuse the ranked list from \textit{BM25} with that of the top-performing dense retriever from our analysis.

\begin{table}[tb]
\centering
\caption{Retrieval models evaluated in the study.}
\label{tab:retrieval_models}
\renewcommand{\arraystretch}{1.2}
\resizebox{\linewidth}{!}{
\begin{tabular}{@{}cccc@{}}
\toprule[1.2pt]
\textbf{Retrieval Paradigm} & \textbf{Model Type} & \textbf{Model Name} & \textbf{Parameter} \\
\midrule
\textbf{Sparse Retrieval} & N/A & BM25 & N/A \\
\midrule
\multirow{6}{*}{\textbf{Dense Retrieval}} & \multirow{4}{*}{General-purpose} & intfloat/multilingual-e5-small (E5)~\cite{wang2024text} & 118M \\
& & ibm-granite/granite-embedding-278m-multilingual (Granite)~\cite{mishra2024granitecodemodelsfamily} & 278M \\
& & Alibaba-NLP/gte-multilingual-base (GTE)~\cite{li2023generaltextembeddingsmultistage} & 305M \\
& & BAAI/bge-m3 (BGE)~\cite{chen2024bgem3embeddingmultilingualmultifunctionality} & 1.3B \\
\cmidrule(lr){2-4}
& \multirow{2}{*}{Code-specialized} & jinaai/jina-embeddings-v2-base-code (Jina)~\cite{gunther2024jinaembeddings28192token} & 161M \\
& & Salesforce/SFR-Embedding-Code-400M\_R (SFR)~\cite{liu2025codexembed} & 434M \\
\bottomrule[1.2pt]
\end{tabular}}
\end{table}

\subsubsection{Context Refinement Stage}
\label{sec:refinement}

This optional stage refines retrieved context to improve the signal-to-noise ratio for the LLM. We systematically evaluate two influential post-retrieval techniques: Re-ranking and Compression, with their effectiveness investigated in RQ3 (\autoref{sec:rq3}).

\textbf{Re-ranking.} This process employs a more powerful but computationally intensive model to re-evaluate the top-$k$ (where $k=20$, a standard threshold balancing computational overhead and recall ceiling) candidates from the initial retrieval, aiming for a more precise relevance ordering~\cite{nogueira2020passagererankingbert}. We contrast two dominant architectural paradigms:

\begin{itemize}
    \item \textbf{Encoder-only:} Conventional cross-encoders concatenate the query and a document, allowing a bidirectional encoder like XLM-RoBERTa to perform deep, token-level interaction~\cite{conneau2020unsupervisedcrosslingualrepresentationlearning}. We use the state-of-the-art \textit{BAAI/bge-reranker-v2-m3}~\cite{chen2024bgem3embeddingmultilingualmultifunctionality}.

    \item \textbf{Decoder-only:} Termed generative re-rankers, these LLMs treat re-ranking as a conditional scoring task~\cite{wang2024large}. We evaluate \textit{Qwen/Qwen3-Reranker-0.6B}~\cite{qwen3embedding}, a specialized model prompted to act as a relevance assessor.
\end{itemize}

\begin{figure}[htbp]
    \centering
    \includegraphics[width=0.75\linewidth]{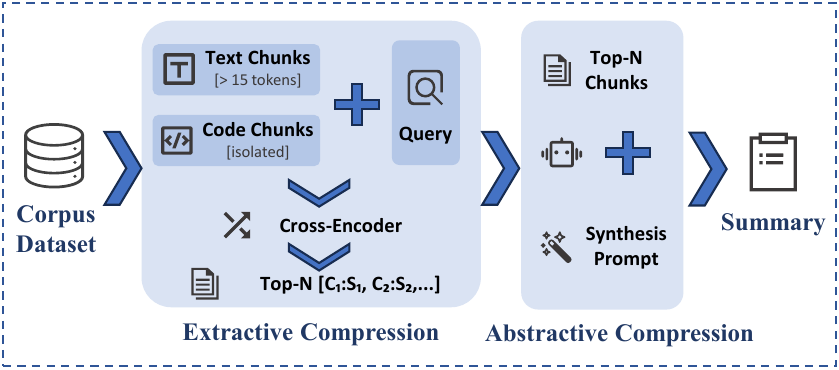}
    \caption{The Extractive-Abstractive Pipeline of the Zero-shot Recomp Adaptation.}
    \Description{The Extractive-Abstractive Pipeline of the Zero-shot Recomp Adaptation.}
    \label{fig:recomp}
\end{figure}

\textbf{Compression.} This technique distills retrieved documents into a more concise format to mitigate issues of limited context windows and noise~\cite{izacard2021leveragingpassageretrieval}. We evaluate two representative query-aware methods:

\begin{itemize}
    \item \textbf{LLMLingua-2:} This technique uses a smaller language model for fine-grained compression~\cite{jiang2024longllmlinguaacceleratingenhancingllms}. It preserves key information by estimating the perplexity of document tokens conditioned on the query, selectively removing those that are least essential.
    
    \item \textbf{Zero-shot Recomp Adaptation:} We adapt the Recomp framework, a two-stage (extractive-abstractive) compression method~\cite{Xu2023RECOMPI}. Our zero-shot version (\autoref{fig:recomp}) avoids the need for task-specific fine-tuning. The workflow is sequential: first, \textbf{Extractive Compression} isolates code blocks and segments text, which is then scored for relevance by a pre-trained cross-encoder. Second, in \textbf{Abstractive Compression}, the top-$N$ chunks are synthesized into a summary by \textit{DeepSeek-V3}. As shown in \autoref{tab:recomp_chunk_stats}, this provides a clear mapping between the number of intermediate chunks ($N$) and their total token count, enabling a direct comparison against the strict token budgets of methods like \textit{LLMLingua-2}.
\end{itemize}

\begin{table}[htbp]
\centering
\caption{Average token count of the top-$N$ chunks after the extractive compression step.}
\label{tab:recomp_chunk_stats}
\renewcommand{\arraystretch}{1.2}
\resizebox{0.55\linewidth}{!}{ 
\begin{tabular}{lcccc}
\toprule[1.2pt]
\textbf{Number of Chunks ($N$)} & 10 & 20 & 40 & 80\\
\midrule
\textbf{Avg. Context Tokens} & 1533.47 & 3578.81 & 7818.43 & 17105.42 \\
\bottomrule[1.2pt]
\end{tabular}}
\end{table}

\subsubsection{Generation Stage}
\label{sec:generation}
In the final stage, a generator synthesizes the query and the context to produce the output. Our evaluation covers a diverse set of highly capable LLMs to represent a spectrum of architectures and sizes. Their impact is evaluated in RQ4 (\autoref{sec:rq4}).

\begin{itemize}
    \item \textbf{Frontier Generalist Models:} To establish the performance ceiling driven by massive reasoning capacity, we evaluate \textit{GPT-4o} \cite{openai2024gpt4o} and \textit{DeepSeek-V3 (671B)} \cite{DeepSeekV2025DeepSeekVTR}.
    
    \item \textbf{Efficient Generalist Models:} To assess the capabilities of widely adopted open-weight models across a diverse range of scales and architectures, we evaluate \textit{Meta Llama 3.3 (70B)} \cite{grattafiori2024llama3herdmodels}, \textit{Qwen2.5-32B-Instruct} (\textit{Qwen2.5}) \cite{qwen2025qwen25technicalreport}, and \textit{Phi-4 (14B)} \cite{abdin2024phi3technicalreporthighly}.

    \item \textbf{Code-Specialized Models:} To explicitly measure the effect of domain-specific training on vast corpora of source code, we evaluate \textit{Qwen2.5-Coder-32B-Instruct} (\textit{Qwen2.5-C}) \cite{qwen2025qwen25technicalreport}.
\end{itemize}

\subsection{LLM-driven Adaptive Configuration}
\label{sec:adaptive}

To translate our empirical findings into a practical solution, we developed a dynamic, LLM-driven adaptive RAG framework. Its core objective is to move beyond static, one-size-fits-all pipelines by autonomously recommending an optimal component configuration for any given SE task. It is built upon a knowledge base of rules and a two-stage reasoning process.

\subsubsection{A Knowledge Base of Empirical Rules}
The framework's foundation is a knowledge base of high-level rules derived from our empirical study. These rules connect observable task features to the performance of specific RAG components across the pipeline's main stages:

\begin{itemize}
    \item \textbf{Query Processing Rules:} Guide the choice of query transformation or augmentation based on the query's initial clarity and detail.
    \item \textbf{Retrieval Rules:} Determine the optimal retrieval paradigm by considering the task's I/O modality (e.g., \textit{text-to-code}, \textit{code-to-text}, or \textit{code-to-code}).
    \item \textbf{Context Refinement Rules:} Govern the activation of optional stages like \textit{Re-ranking} and \textit{Compression} based on factors like query ambiguity and context window.
    \item \textbf{Generation Rules:} Provide guidance on selecting a generator by weighing the trade-offs between generalist and code-specialized models based on task complexity.
\end{itemize}

\subsubsection{Two-Stage Adaptive Process}
\begin{figure}[htbp]
    \centering
    \includegraphics[width=\linewidth]{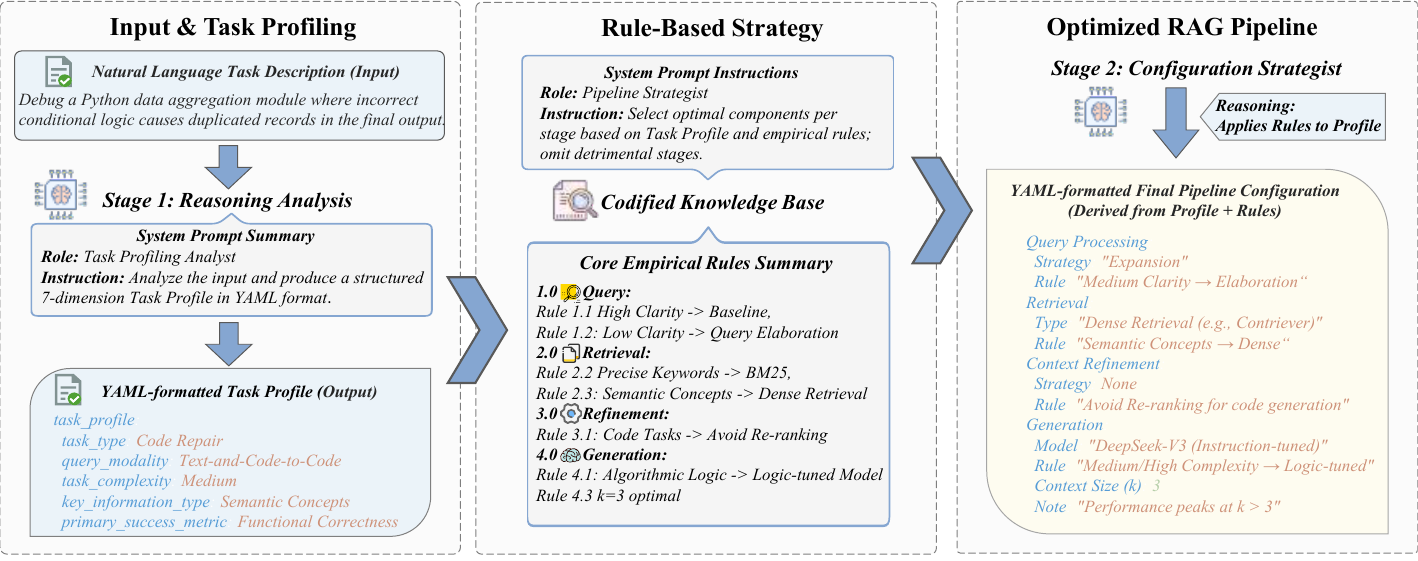}
    \Description{The Decision Logic of the LLM-Driven Adaptive RAG Framework.}
    \caption{The Decision Logic of the LLM-Driven Adaptive RAG Framework.}
    \label{fig:prompt_design}
\end{figure}

To apply this knowledge base automatically, we developed a prototype adaptive framework. As visualized in \autoref{fig:prompt_design}, this framework employs a dynamic, two-stage inference process powered by a large language model (\textit{DeepSeek-V3}).

\noindent\textbf{Stage 1: Automated Task Profiling.} The framework parses the natural language description of an unseen SE task to generate a structured \textit{Task Profile}. It evaluates the input across critical dimensions (query modality, clarity, complexity, and key information type) to accurately capture the task's semantic and structural demands.

\noindent\textbf{Stage 2: Knowledge-Driven Configuration.} The extracted profile is dynamically mapped against our codified \textit{Knowledge Base of Empirical Rules}. Governed by the decision trees in \autoref{fig:prompt_design}, the framework bypasses static defaults to autonomously recommend the optimal pipeline.

By grounding its decisions entirely in our empirical evidence, the framework ensures strict theoretical alignment. The external validation of this framework's generalization capabilities is presented in \autoref{sec:adaptive_validation}.

\section{Evaluation}
\label{sec:evaluation}

\subsection{Datasets and Corpora}
\label{sec:datasets and corpora}

All experiments focus on the Python language to ensure a controlled environment, leveraging its widespread adoption and the abundance of high-quality public data.

\subsubsection{Task Datasets}
\label{sec:datasets}
We selected three datasets for our evaluation suite, each tailored to a core code-related task and preprocessed to create focused, balanced evaluation sets. 

\begin{itemize}    
    \item \textbf{Code Generation:} We employ the APPS benchmark \cite{hendrycks2021measuringMathematicalProblem}, constructing a balanced 300-problem subset by randomly sampling 100 entries from each of its three difficulty tiers.

    \item \textbf{Code Summarization:} Using the Python subset of CodeXGLUE~\cite{lu2021codexglue}, we explicitly select the 100 longest snippets (Avg: $\sim$1,510 tokens vs. Global: $\sim$206) to mitigate ceiling effects and stress-test RAG capabilities. To isolate syntactic reliance, we created an obfuscated version (CodeXGLUE-OBF): using Python's AST, we popped docstrings and deterministically mapped four user identifier types (functions, classes, arguments, variables) to sequential placeholders (e.g., \texttt{var\_1}), while strictly preserving built-ins, imported modules, attributes, and magic methods.

    \item \textbf{Code Repair:} We utilize DebugBench~\cite{tian2024debugbenchevaluatingdebuggingcapability}, curating a 300-sample test set by selecting 25 samples from each of the 12 difficulty-category combinations (covering \textit{syntax}, \textit{reference}, \textit{logic}, and \textit{multiple} bug types).
\end{itemize}

\subsubsection{Retrieval Corpora}
\label{sec:corpora}
To construct a comprehensive knowledge base for our RAG system, we aggregated and processed data from five diverse sources, summarized in \autoref{tab:corpora_summary}.

\begin{table}[htbp]
\centering
\caption{Document Counts of the Retrieval Corpora}
\label{tab:corpora_summary}
\renewcommand{\arraystretch}{1.2}
\resizebox{0.9\linewidth}{!}{
\begin{tabular}{@{}ccccccc@{}}
\toprule[1.2pt]
\textbf{Corpus} & \textbf{Stack Overflow} & \textbf{Python API} & \textbf{LeetCode} & \textbf{CodeSearchNet} & \textbf{Code-Contests} & \textbf{Total} \\
\midrule
\textbf{Documents} & 164,085 & 38,352 & 3,174 & 13,590 & 13,432 & \textbf{232,633} \\
\bottomrule[1.2pt]
\end{tabular}}
\end{table}

The foundation of the corpus combines community-driven knowledge with canonical documentation via strict quantitative filtering. For \textbf{Stack Overflow}, we extracted Python-related posts by enforcing quality thresholds (Question Score $\ge 10$, Answer Score $\ge 5$), retaining only the top-5 answers truncated to 512 tokens. This is complemented by \textbf{Python API} entries, programmatically parsed from 33 widely adopted industry libraries (e.g., \texttt{numpy}, \texttt{pandas}) to serve as functional references. We curated \textbf{LeetCode} by pairing algorithmic problems with 1-2 high-quality reference solutions, alongside deduplicated function-docstring pairs from \textbf{CodeSearchNet}. Finally, we processed \textbf{Code-Contests} by filtering strictly for Python solutions and flattening the structure, retaining a maximum of 10 unique problem-solution entries per problem.

As performance generally peaks at $k=3$ (\autoref{sec:discussion_k_value}), we standardize this retrieval depth across RQ1-RQ3 to strictly isolate the impact of individual RAG components. Additionally, we validate the necessity of our heterogeneous corpus via source ablation (\autoref{sec:source_ablation}).

\subsection{Experimental Setup}
\label{sec:experimental_setup}

\subsubsection{Evaluation Metrics}
\label{sec:evaluation_metrics}
We employ metrics tailored to each task's unique objectives, citing their foundational sources to ensure methodological rigor.

For \textbf{Code Generation} and \textbf{Code Repair}, we evaluate functional correctness and structural quality through the following metrics:

\begin{itemize}
    \item \textbf{Weighted Pass@1 (W-Pass@1)} is based on the \textit{Pass@k}~\cite{chen2021evaluating} which estimates the probability that at least one of $k$ generated samples is correct.
    \begin{equation}
    \text{Pass@k} = \mathbb{E}_{\text{Problems}} \left[ 1 - \frac{\binom{n-c}{k}}{\binom{n}{k}} \right]
    \end{equation}
    where $n$ is the samples per problem and $c$ is the number of correct ones. For our experiments ($k=1$), we calculate a weighted average to account for problem difficulty. Informed by prior work on difficulty-based sample weighting~\cite{zhou2022understanding}, we assign a higher weight to more challenging problems to better reflect their significance. The final metric is:
    \begin{equation}
    \text{W-Pass@1} = \sum_{p \in P} w_p \cdot \text{Pass@1}_p \quad \text{where} \quad w_p = \frac{d_p}{\sum_{p' \in P} d_{p'}}
    \end{equation}

    \item \textbf{CodeBLEU} serves as a structure-aware metric for \textbf{Code Repair}. It complements binary correctness by comparing a solution's syntactic (AST) and data-flow similarity to the reference~\cite{ren2020codebleu}. The metric is a weighted combination of these components:
    \begin{equation}
    \text{CodeBLEU} = \alpha \cdot \text{BLEU}_{\text{weighted}} + \beta \cdot \text{BLEU}_{\text{ngram}} + \gamma \cdot \text{Match}_{\text{AST}} + \delta \cdot \text{Match}_{\text{DF}}
    \end{equation}
\end{itemize}

For \textbf{Code Summarization}, we evaluate the semantic alignment with the ground truth.
\begin{itemize}
    \item \textbf{Semantic Similarity (Sim$_{\text{Emb}}$)} computes the cosine similarity between the vector embeddings of the candidate and reference summaries. We generate embeddings using the \textit{intfloat/e5-mistral-7b-instruct} model~\cite{wang2024text}.
    \begin{equation}
    \text{Sim}_\text{Emb}(c, r) = \frac{\mathbf{v}_{c} \cdot \mathbf{v}_{r}}{\|\mathbf{v}_{c}\| \|\mathbf{v}_{r}\|}
    \end{equation}
    where $\mathbf{v}_{c}$ and $\mathbf{v}_{r}$ are the corresponding vector embeddings.
\end{itemize}

\subsubsection{Experimental Environment}
\label{sec:experimental_environment}
The experiments were conducted on a server equipped with an NVIDIA A100 GPU with 80GB of memory. The server operates under a Linux environment, specifically, Linux version 5.15.0-97-generic, compiled using GCC (Ubuntu 11.4.0-1ubuntu1~22.04) version 11.4.0 and GNU ld (GNU Binutils for Ubuntu) version 2.38. 

The various models in our testbed were deployed as follows: \textit{DeepSeek-V3} and \textit{Qwen2.5} were accessed via the SiliconFlow API, while other generators, such as \textit{Llama-3.3-70B} and \textit{Phi-4-14B}, were run locally using the Ollama framework. All models for the intermediate RAG stages—including dense retrievers, rerankers (\textit{e.g., BAAI/bge-reranker-v2-m3}), and compressors—were hosted and executed within our local environment.

\subsection{RQ1: Impact of the \textit{Query Processing Stage}}
\label{sec:rq1}

We investigate the impact of query refinement by evaluating the four strategies detailed in \autoref{sec:query_processing}: \textbf{Baseline} (the original query), \textbf{Simplified}, \textbf{Elaborated}, and the oracle-augmented \textbf{HyDE}. The primary analysis, summarized in \autoref{tab:rq1_query}, focuses on performance when retrieving the top-3 documents ($k=3$).

\begin{table}[htbp]
\centering
\caption{Performance comparison of Query Processing strategies at $k=3$. Parentheses denote relative percentage changes vs. \textbf{Baseline}. Best results in \textbf{bold}. (CXG: CodeXGLUE, DB: DebugBench, O: Obfuscated).}
\label{tab:rq1_query}
\resizebox{\linewidth}{!}{
\begin{tabular}{ll c cc cc c}
\toprule[1.2pt]
\multirow{2}{*}{\textbf{Task}} & \multirow{2}{*}{\textbf{Dataset (Metric)}} & \multirow{2}{*}{\textbf{Baseline}} & \multicolumn{2}{c}{\textbf{Simplified}} & \multicolumn{2}{c}{\textbf{Elaborated}} & \multirow{2}{*}{\textbf{HyDE}} \\
\cmidrule(lr){4-5} \cmidrule(lr){6-7}
& & & DeepSeek-V3 & GPT-4o & DeepSeek-V3 & GPT-4o \\
\midrule
Code Gen. & APPS (\textit{W-Pass@1}) & \textbf{38.36} & 27.91 (-27.2\%) & 25.81 (-32.7\%) & 33.72 (-12.1\%) & 22.04 (-42.5\%) & 31.45 (-18.0\%) \\
\midrule
\multirow{2}{*}{Code Sum.} & CXG (\textit{Sim$_{\text{Emb}}$}) & 87.75 & 87.78 (+0.03\%) & 87.50 (-0.28\%) & \textbf{87.84 (+0.10\%)} & 87.61 (-0.16\%) & 87.71 (-0.05\%) \\
& CXG-O (\textit{Sim$_{\text{Emb}}$}) & 74.50 & 80.21 (+7.66\%) & 79.83 (+7.15\%) & \textbf{80.38 (+7.89\%)} & 80.11 (+7.53\%) & 75.05 (+0.74\%) \\
\midrule
\multirow{2}{*}{Code Repair} & DB (\textit{W-Pass@1}) & 76.24 & 69.44 (-8.92\%) & 75.45 (-1.04\%) & 73.08 (-4.14\%) & \textbf{77.35 (+1.46\%)} & 75.88 (-0.47\%) \\
& DB (\textit{CodeBLEU}) & 46.32 & \textbf{56.65 (+22.3\%)} & 51.93 (+12.1\%) & 47.74 (+3.07\%) & 49.87 (+7.66\%) & 47.24 (+1.99\%) \\
\bottomrule[1.2pt]
\end{tabular}}
\end{table}

\subsubsection{Cross-Task Performance Analysis}
The effectiveness of query processing is highly task-dependent, a phenomenon observed consistently across both transformation models.

For the APPS code generation task, which characterizes \textit{well-defined tasks} featuring explicit problem statements, strict I/O examples, and clear constraints, the unaltered \textbf{Baseline} query is superior (\textit{W-Pass@1} 38.36). All LLM-based transformations severely degrade performance, with \textbf{Simplified} and \textbf{Elaborated} queries causing drops up to 32.7\% and 42.5\% across both models. This suggests for tasks with explicit requirements and robust examples, automated transformations may strip away critical lexical details essential for exact-match retrieval.

Conversely, query transformation is highly beneficial for \textit{ambiguous or noisy inputs}, such as obfuscated identifiers or informal, incomplete user queries. On the CodeXGLUE-OBF dataset, the \textbf{Elaborated} strategy improves \textit{Sim$_{\text{Emb}}$} by 7.89\% with \textit{DeepSeek-V3} and 7.53\% with \textit{GPT-4o}. Without explicit semantic identifiers, elaboration forces the LLM to hypothesize the underlying structural intent, generating useful contextual keywords for retrieval. However, on the standard CodeXGLUE dataset, the same method offers a negligible 0.1\% gain at most. This indicates that while query transformation is directionally helpful for noisy code, its impact is far more constrained than in natural language contexts \cite{Ma2023QueryRewriting}.

DebugBench results highlight a critical trade-off. The \textbf{Simplified} query dramatically improves the \textit{CodeBLEU} score (up to +22.3\%), indicating better structural similarity to the reference. However, this gain sacrifices functional correctness (\textit{W-Pass@1} declines of 8.92\% and 1.04\%), suggesting that simplification may prioritize superficial structure over function.

\subsubsection{Findings of RQ1}
Based on this analysis, we derive the following findings:
\begin{itemize}
    \item \textbf{Finding 1:} The original, unaltered query is a powerful baseline, often outperforming automated transformations on tasks with clear, unambiguous inputs.
    \item \textbf{Finding 2:} Query elaboration is more effective for ambiguous or noisy inputs, where it can provide crucial context and significantly improve retrieval relevance.
    \item \textbf{Finding 3:} Query simplification can trade functional correctness for structural similarity.
    \item \textbf{Finding 4:} All LLM-based query transformations risk the unintentional alteration of a query's core intent or the omission of critical details.
\end{itemize}

\find{\textbf{Answer to RQ1}:
The impact of the Query Processing stage is highly context-dependent. For well-defined tasks, the original query is most robust. For ambiguous tasks, query elaboration can provide significant benefits, but automated transformations should be applied cautiously due to the risk of semantic drift, regardless of the LLM used.}

\subsection{RQ2: Comparative Analysis of the \textit{Retrieval Stage}}
\label{sec:rq2}

In this section, we evaluate the core of the RAG pipeline by comparing the three retrieval paradigms detailed in \autoref{sec:retrieval}: \textbf{Sparse Retrieval}, \textbf{Dense Retrieval}, and \textbf{Hybrid Retrieval}. We benchmark their performance against a non-RAG \textit{Zero-Shot} baseline and an \textit{Oracle} setup, representing the theoretical upper bound. The detailed results are presented in \autoref{tab:rq2_retrival}.

\begin{table}[htbp]
\centering
\caption{Comparative performance of Retrieval strategies at $k=3$. Parentheses denote relative percentage changes compared to the \textbf{Zero-Shot} baseline. Best results for each metric are highlighted in \textbf{bold}.}
\label{tab:rq2_retrival}
\resizebox{\linewidth}{!}{%
\begin{tabular}{@{}lccccc@{}}
\toprule[1.2pt]
& \textbf{Code Generation} & \multicolumn{2}{c}{\textbf{Code Summarization}} & \multicolumn{2}{c}{\textbf{Code Repair}} \\
\cmidrule(lr){2-2} \cmidrule(lr){3-4} \cmidrule(lr){5-6}
\textbf{Retrieval Paradigm} & APPS & CodeXGLUE & CodeXGLUE-OBF & \multicolumn{2}{c}{DebugBench} \\
& (\textit{W-Pass@1}) & (\textit{Sim$_{\text{Emb}}$}) & (\textit{Sim$_{\text{Emb}}$}) & (\textit{W-Pass@1}) & (\textit{CodeBLEU}) \\
\midrule
\multicolumn{6}{@{}l}{\textit{Baselines}} \\
\quad Zero-Shot & 15.18 & 87.38 & 75.24 & 64.03 & 66.26 \\
\quad Oracle & 55.99 (+268.8\%) & 88.28 (+1.0\%) & 86.47 (+15.0\%) & 70.01 (+9.3\%) & 75.28 (+13.6\%) \\
\midrule
\multicolumn{6}{@{}l}{\textit{Sparse Retrieval}} \\
\quad BM25 & \textbf{38.00 (+150.3\%)} & \textbf{87.90 (+0.6\%)} & \textbf{75.61 (+0.5\%)} & \textbf{76.53 (+19.5\%)} & 46.39 (-30.0\%) \\
\midrule
\multicolumn{6}{@{}l}{\textit{Dense Retrieval (General-purpose)}} \\
\quad E5 (118M) & 33.72 (+122.1\%) & 87.86 (+0.5\%) & 74.82 (-0.6\%) & 73.70 (+15.1\%) & 46.38 (-30.0\%) \\
\quad Granite (278M) & 35.63 (+134.7\%) & 87.73 (+0.4\%) & 75.35 (+0.1\%) & 74.44 (+16.3\%) & 46.91 (-29.2\%) \\
\quad GTE (305M) & 34.45 (+126.9\%) & \textbf{87.90 (+0.6\%)} & 75.05 (-0.3\%) & 76.37 (+19.3\%) & 44.87 (-32.3\%) \\
\quad BGE (1.3B) & 25.82 (+70.1\%) & 87.61 (+0.3\%) & 74.67 (-0.8\%) & 68.13 (+6.4\%) & \textbf{63.57 (-4.1\%)} \\
\midrule
\multicolumn{6}{@{}l}{\textit{Dense Retrieval (Code-specialized)}} \\
\quad Jina (161M) & 27.09 (+78.4\%) & 87.67 (+0.3\%) & 74.93 (-0.4\%) & 68.49 (+7.0\%) & 59.73 (-9.8\%) \\
\quad SFR (434M) & 34.27 (+125.7\%) & 87.79 (+0.5\%) & 74.56 (-0.9\%) & 72.84 (+13.8\%) & 47.60 (-28.2\%) \\
\midrule
\multicolumn{6}{@{}l}{\textit{Hybrid Retrieval}} \\
\quad Hybrid & 33.54 (+120.9\%) & 87.68 (+0.3\%) & 75.22 (-0.0\%) & 75.04 (+17.2\%) & 47.37 (-28.5\%) \\
\bottomrule[1.2pt]
\end{tabular}}
\end{table}

\subsubsection{Cross-Task Performance Analysis}
The results unequivocally show that all retrieval paradigms substantially outperform the \textit{Zero-Shot} baseline, confirming the fundamental value of RAG for code tasks. However, the optimal strategy is highly task-dependent.

The most striking result is the exceptional performance of \textit{BM25}, a classic \textbf{Sparse Retrieval} method. Unlike general-domain NLP tasks where dense retrievers typically dominate \cite{Karpukhin2020DensePR, gao2024retrievalaugmented}, SE tasks often impose hard lexical constraints. On the APPS task, \textit{BM25}'s \textit{W-Pass@1} of 38.00 (+150.3\%) highlights that precise keyword matching of algorithms is crucial; dense models may suffer from semantic drift, retrieving conceptually similar but syntactically incompatible code. This exact-match advantage persists on code-based queries: \textit{BM25} achieves the highest \textit{Sim$_{\text{Emb}}$} on standard CodeXGLUE (87.90) and noisy CodeXGLUE-OBF (75.61), suggesting it robustly captures structural control-flow keywords even when semantic identifiers vanish. Finally, it leads in functional correctness on DebugBench with a \textit{W-Pass@1} of 76.53, likely by matching exact error traces.

\textbf{Dense Retrieval}, while powerful, exhibits more nuanced performance. Our findings reveal two key insights. First, model scale is not a decisive factor; the smaller, 305M-parameter \textit{GTE} model consistently outperforms the largest 1.3B-parameter \textit{BGE} model, particularly on APPS (34.45 vs. 25.82 \textit{W-Pass@1}). Second, code-specialization does not guarantee an advantage, as the code-trained \textit{Jina} and \textit{SFR} models failed to surpass the best general-purpose models like \textit{GTE}.

\textbf{Hybrid Retrieval} proves to be a reliable but conservative strategy, delivering consistent middle-of-the-pack results but failing to unlock any significant synergistic gains. Finally, the DebugBench results highlight a critical trade-off: RAG methods consistently improve functional correctness at the cost of structural similarity. For instance, while \textit{BM25} boosts \textit{W-Pass@1} by +19.5\%, it degrades the \textit{CodeBLEU} score by 30.0\%, suggesting that retrieving externally correct but different code patterns steers the generator away from minimal, structurally-similar edits.

\subsubsection{Findings of RQ2}
\begin{itemize}
    \item \textbf{Finding 5:} RAG is fundamentally beneficial for code tasks, as all evaluated retrieval paradigms significantly outperform the non-retrieval \textit{Zero-Shot} baseline.
    \item \textbf{Finding 6:} The lexical-based \textit{BM25} is an exceptionally powerful and versatile retriever, excelling across text-heavy, clean code, and noisy code queries via precise term matching.
    \item \textbf{Finding 7:} Model scale is not a decisive factor in dense retrieval performance. We found that smaller, efficient models like \textit{GTE} consistently surpassed the largest model, \textit{BGE}.
    \item \textbf{Finding 8:} Domain specialization on code does not guarantee superior performance, as the best general-purpose retrievers were unbeaten even on code-centric tasks.
    \item \textbf{Finding 9:} \textit{Hybrid Retrieval} is a safe but conservative strategy that, in our experiments, did not yield performance gains beyond the single best retriever for a given task.
\end{itemize}

\find{\textbf{Answer to RQ2}:
No single retrieval strategy is universally superior, as the optimal choice depends on the task's query modality. However, our findings strongly suggest that a dual-retriever approach, leveraging the exceptional lexical precision of \textit{BM25} alongside the semantic understanding of a robust general-purpose dense retriever like \textit{GTE}, provides the most consistent and high-performing foundation for a wide array of code-related tasks.}

\subsection{RQ3: Contribution of the \textit{Context Refinement Stage}}
\label{sec:rq3}

In this section, we evaluate the two primary techniques of the Context Refinement stage as defined in \autoref{sec:refinement}: \textbf{Re-ranking} and \textbf{Compression}. Our objective is to determine if these post-retrieval steps improve the signal-to-noise ratio and enhance final task performance.

\subsubsection{Re-ranking Performance Analysis}
We analyzed the impact of re-ranking by applying two state-of-the-art models to the outputs of three representative retrievers: the top sparse model (\textit{BM25}), an efficient dense model (\textit{GTE}), and a weaker dense model (\textit{BGE}). As shown in \autoref{tab:rq3_reranker}, our findings indicate that re-ranking is an unstable and often detrimental step for code-related tasks.

\begin{table}[htbp]
\centering
\caption{Performance of Reranker-Retriever combinations at $k=3$. Parentheses denote relative percentage changes against the corresponding un-reranked baselines (\autoref{tab:rq2_retrival}). Best results are highlighted in \textbf{bold}.}
\label{tab:rq3_reranker}
\resizebox{\linewidth}{!}{
\begin{tabular}{lcccccc}
\toprule[1.2pt]
& & \textbf{Code Generation} & \multicolumn{2}{c}{\textbf{Code Summarization}} & \multicolumn{2}{c}{\textbf{Code Repair}} \\
\cmidrule(lr){3-3} \cmidrule(lr){4-5} \cmidrule(lr){6-7}
\textbf{Reranker} & \textbf{Base Retriever} & APPS & CodeXGLUE & CodeXGLUE-OBF & DebugBench & DebugBench \\
& & (\textit{W-Pass@1}) & (\textit{Sim$_{\text{Emb}}$}) & (\textit{Sim$_{\text{Emb}}$}) & (\textit{W-Pass@1}) & (\textit{CodeBLEU}) \\
\midrule
\multirow{3}{*}{BEG-Reranker} & BM25 & 33.36 (-12.21\%) & 87.87 (-0.03\%) & 74.72 (-1.18\%) & \textbf{76.76 (+0.30\%)} & 46.74 (+0.75\%) \\
& GTE  & 32.72 (-5.02\%) & 87.44 (-0.51\%) & 75.00 (-0.07\%) & 75.55 (-1.07\%) & 47.73 (+6.37\%) \\
& BGE  & 23.82 (-7.75\%) & 87.36 (-0.29\%) & 74.73 (+0.08\%) & 67.48 (-0.95\%) & 62.92 (-1.02\%) \\
\midrule
\multirow{3}{*}{Qwen-Reranker} & BM25 & 34.72 (-8.63\%) & \textbf{87.90 (0.00\%)} & 75.59 (-0.03\%) & 72.32 (-5.50\%) & 47.79 (+3.02\%) \\
& GTE  & 34.09 (-1.04\%) & 87.68 (-0.24\%) & 75.04 (-0.01\%) & 73.27 (-4.06\%) & 48.23 (+7.49\%) \\
& BGE  & 23.36 (-9.53\%) & 87.52 (-0.10\%) & 74.67 (0.00\%) & 68.74 (+0.90\%) & 62.91 (-1.04\%) \\
\bottomrule[1.2pt]
\end{tabular}}
\end{table}

For Code Generation, re-ranking consistently harms functional correctness. \textit{W-Pass@1} scores drop substantially in all configurations; the best retriever, \textit{BM25}, declines by up to 12.21\%. Notably, re-ranking also fails to salvage weaker retrievals from \textit{BGE}, further degrading its score. For Code Summarization, the impact is negligible, failing to justify the added computational cost.

DebugBench results are particularly unreliable. While \textit{BGE-Reranker} offers a marginal +0.30\% \textit{W-Pass@1} gain for \textit{BM25}, this lone outlier contrasts with a 5.50\% correctness drop when paired with \textit{Qwen-Reranker}. Although some configurations slightly improve structural similarity (\textit{CodeBLEU}), these minor gains cannot offset the significant, unpredictable degradation in functional correctness.

\subsubsection{Compression Performance Analysis}

\begin{figure}[htbp]
    \centering
    \includegraphics[width=1\textwidth]{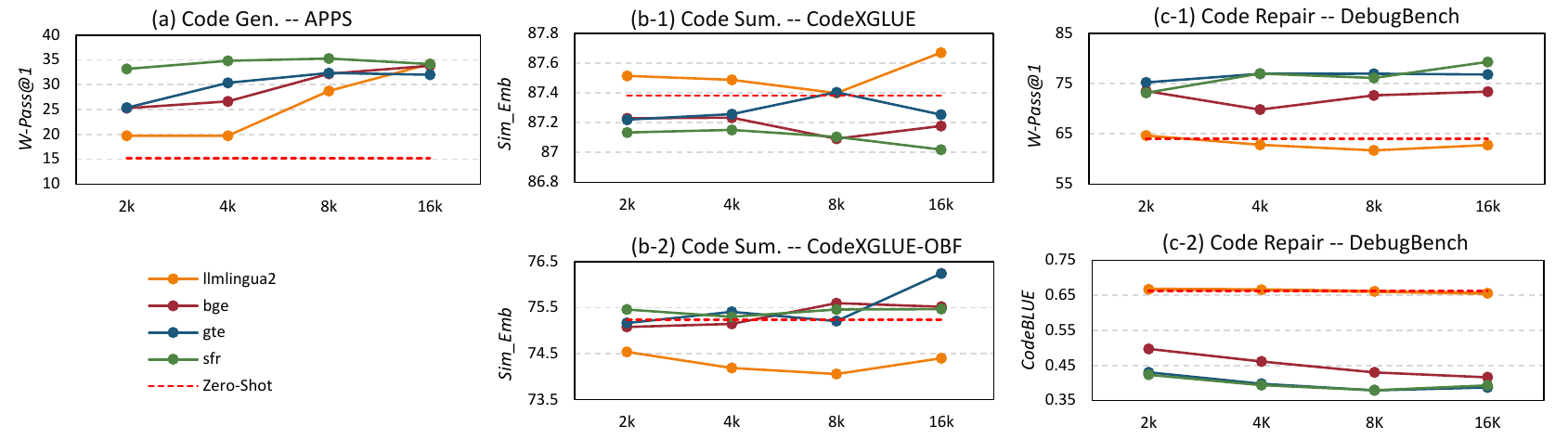}
    \caption{Performance of context compression methods across five evaluation scenarios. The charts illustrate how performance (y-axis) changes with an increasing token budget (x-axis) for each compressor. The dashed line represents the \textit{Zero-Shot} baseline.}
    \Description{Performance of context compression methods across five evaluation scenarios. The charts illustrate how performance (y-axis) changes with an increasing token budget (x-axis) for each compressor. The dashed line represents the \textit{Zero-Shot} baseline.}
    \label{fig:compression_results}
\end{figure}

We evaluate context compression by comparing two distinct methodologies: the token-level pruning of \textit{LLMLingua-2} and variants of our \textit{Zero-Shot Recomp Adaptation} which use different cross-encoders (\textit{GTE}, \textit{BGE}, and the code-specialized \textit{SFR}). As illustrated in \autoref{fig:compression_results}, we benchmarked these against a \textit{Zero-Shot} floor and an uncompressed \textit{Optimal Retriever} ceiling.

For APPS code generation, all compressors improve performance as the token budget increases. The code-specialized \textit{SFR} adaptation proves most effective, peaking around an 8000-token budget. Conversely, compression is largely ineffective for Code Summarization. On standard CodeXGLUE, all methods trail the uncompressed retriever; even on the obfuscated version, gains remain minor, suggesting the task is too nuance-sensitive for current techniques.

DebugBench reveals a complex trade-off. For functional correctness (\textit{W-Pass@1}), the \textit{SFR} and \textit{GTE} adaptations show a remarkable upward trend, surpassing both the \textit{Zero-Shot} baseline and the uncompressed \textit{Optimal Retriever}, suggesting compression effectively filters noise. However, the \textit{CodeBLEU} chart tells the opposite story: all curves trend downwards as token budgets increase. This confirms a critical trade-off: compression achieves functional correctness by progressively discarding original stylistic and structural context.

\subsubsection{Findings of RQ3}
Our analysis of the context refinement stage leads to the following findings:
\begin{itemize}
    \item \textbf{Finding 10:} Re-ranking is an unreliable component that fails to justify its computational overhead. It often degrades the performance of strong, keyword-based retrievals by disrupting an already optimal ordering.
    \item \textbf{Finding 11:} The corrective potential of re-ranking is minimal. It fails to consistently or substantially improve results from weaker initial retrievals, making it an ineffective tool for salvaging performance.
    \item \textbf{Finding 12:} The utility of context compression is critically dependent on the token budget and the choice of compressor. Code-specialized models with a sufficient budget are more robust, while aggressive compression is consistently harmful.
\end{itemize}

\find{\textbf{Answer to RQ3}:
The Context Refinement stage is not a guaranteed improvement and should be applied with caution. Our findings show that \textbf{Re-ranking} is generally an inadvisable step due to its high cost and unreliable, often negative, impact. \textbf{Compression} reserved for models with strict context length limitations. If used, a high token budget ($ \ge 8000$) and a compatible compressor are critical to mitigate performance loss. Ultimately, employing a generator with a native long-context window is a superior strategy to relying on this stage.} 

\subsection{RQ4: Influence of the \textit{Generator Stage}}
\label{sec:rq4}
\begin{figure}[!ht]
    \centering
    \includegraphics[width=1\textwidth]{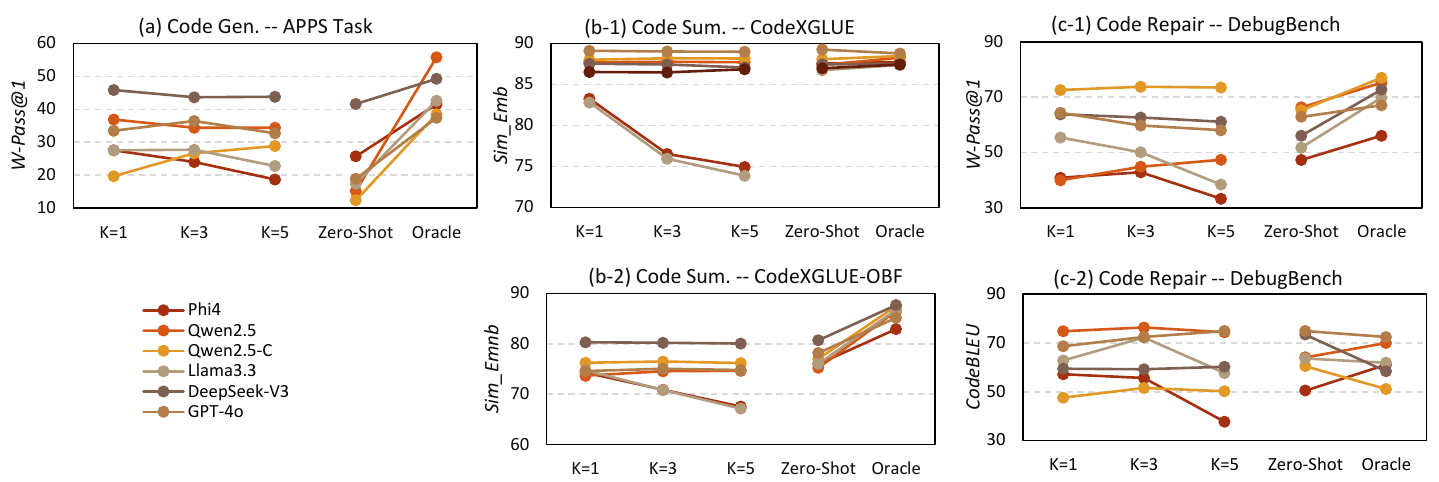}
    \caption{Performance comparison of different generators across all tasks. RAG performance is shown for $k=1, 3, 5$ retrieved documents, with \textit{Zero-Shot} and \textit{Oracle} results as baselines.}
    \Description{Performance comparison of different generators across all tasks. RAG performance is shown for $k=1, 3, 5$ retrieved documents, with \textit{Zero-Shot} and \textit{Oracle} results as baselines.}
    \label{fig:generation_results}
\end{figure}

In this section, we evaluate the impact of the generator (\autoref{sec:generation}), analyzing how the capabilities of different LLMs establish the performance ceiling of the RAG system, as illustrated in \autoref{fig:generation_results}.

\subsubsection{Cross-Task Performance Analysis}
Universally, all evaluated models benefit from retrieval, with RAG-based scores at $k=1, 3, 5$ consistently outperforming their respective \textit{Zero-Shot} baselines. \textit{Oracle} context performance further confirms the profound impact of context quality on generation.

On the APPS code generation task, \textit{DeepSeek-V3} is the clear frontrunner, alongside highly competitive performance from \textit{GPT-4o}. Both frontier models demonstrate superior problem-solving capabilities via distinctly higher performance curves (\autoref{fig:generation_results}(a)). For most models, performance peaks at $k=3$ before plateauing or declining, suggesting that a small number of high-quality documents is optimal while excess context introduces noise.

On the standard CodeXGLUE dataset, top models including \textit{GPT-4o}, \textit{Qwen2.5-C}, and \textit{DeepSeek-V3} are tightly clustered near the ceiling. However, the challenging obfuscated dataset reveals clearer model strengths. Here, the raw reasoning capacity of \textit{DeepSeek-V3} and the domain specialization of \textit{Qwen2.5-C} become apparent. This validates the distinct benefits of massive model scale and code-specific training on structurally ambiguous inputs.

The DebugBench repair task highlights clear task-specific advantages. While \textit{DeepSeek-V3} dominated generation, \textit{Qwen2.5} and \textit{GPT-4o} prove highly effective for code repair, achieving the highest functional correctness (\textit{W-Pass@1}). Crucially, \textit{CodeBLEU} results (\autoref{fig:generation_results}(c-2)) confirm a recurring finding: for most generators, RAG improves functional correctness at the cost of structural similarity when compared to the \textit{Zero-Shot} baseline.

\subsubsection{Findings of RQ4}
Our analysis of the generator stage leads to the following findings:
\begin{itemize}
    \item \textbf{Finding 13:} A generator's effectiveness depends on both its intrinsic reasoning capability (standalone performance) and its ability to synthesize retrieved context.
    \item \textbf{Finding 14:} Model strengths are highly task-dependent. Frontier models like \textit{DeepSeek-V3} and \textit{GPT-4o} excel at complex generative reasoning, while specialized models like \textit{Qwen2.5-C} are highly effective for structural code comprehension.
    \item \textbf{Finding 15:} More retrieved context is not always better. For complex tasks like code generation, performance often peaks with a small number of documents ($k=3$), as additional context can introduce noise.
    \item \textbf{Finding 16:} The trade-off between functional correctness and structural similarity in code repair is a consistent phenomenon observed across most generators when using RAG.
\end{itemize}

\find{\textbf{Answer to RQ4}:
The generator is a crucial component whose capabilities establish the performance ceiling for a RAG system. The optimal choice is task-dependent, requiring a balance between the model's intrinsic ability and its capacity to leverage context. Effective retrieval augmentation is therefore not just about selecting a powerful generator, but also about providing it with the optimal amount of high-quality information for the specific task at hand.}

\section{Discussion}
\label{sec:discussion}

\subsection{Summary of Key Findings and Implications}
\label{sec:discussion_findings}

Our empirical study on RAG for SE reveals several critical insights. Our findings point not to a simple hierarchy of components, but to a complex interplay between task profiles and RAG pipeline design. We synthesize our key findings and their implications below:

\begin{itemize}
    \item \textbf{The Primacy of the Retriever:} Our results show that the initial retrieval stage is the most critical determinant of pipeline performance. Strikingly, the classic lexical retriever, \textit{BM25}, proved exceptionally effective, frequently outperforming larger, dense models across diverse query modalities (\autoref{sec:rq2}). Its success in matching specific algorithm names and error messages underscores the enduring importance of lexical precision in code tasks (\textbf{Finding 6}). For practitioners, this implies that prioritizing a robust and suitable retriever is often a more impactful investment than focusing solely on the generator's scale.
    
    \item \textbf{The Situational Value of Advanced Components:} Our findings caution against building unnecessarily complex pipelines, as the value of advanced components is highly situational and their indiscriminate use is often detrimental. While query elaboration benefited ambiguous inputs (\textbf{Finding 2}), most transformations introduced semantic drift on well-defined tasks (\textbf{Finding 4}). Similarly, re-ranking failed to correct weak retrievals and actively harmed the ordering of strong ones (\textbf{Finding 10}). Practitioners should therefore view these components not as default enhancements, but as specialized tools for targeted problems.
    
    \item \textbf{No One-Size-Fits-All:} A central conclusion is that no single RAG configuration is universally optimal. The ideal component choice is coupled with the task's nature, as the best retrieval strategy (\autoref{sec:rq2}), query method (\autoref{sec:rq1}), and generator (\autoref{sec:rq4}) all varied significantly. For instance, \textit{DeepSeek-V3} excelled at code generation while \textit{Qwen2.5-C} is more effective for code comprehension (\textbf{Finding 14}). This highlights a critical implication for researchers and tool builders: the future of Code RAG lies in moving beyond static pipelines towards adaptive frameworks that dynamically configure the optimal component chain.
\end{itemize}

\subsection{Validation of the Adaptive Framework}
\label{sec:adaptive_validation}

As introduced in our approach (\autoref{sec:adaptive}), we developed and open-sourced an adaptive RAG framework. To validate the framework's generalization beyond initial datasets, we conducted simulations using problem descriptions from three external SE studies: \textit{Move Method Refactoring}~\cite{Bellur2025ICSME}, \textit{Test Generation}~\cite{shin2024retrievalaugmentedtestgenerationfar}, and \textit{Library Code Generation}~\cite{zhou2023docprompting}. 

For each case, the framework autonomously profiled the task and recommended architectural choices perfectly aligning with the original authors' manual best practices (\autoref{tab:external_validation}). Crucially, its internal reasoning dynamically synthesizes our empirical findings across the entire RAG pipeline:

\begin{itemize}
    \item \textbf{Bridging Semantic Gaps:} For \textit{Move Method Refactoring}~\cite{Bellur2025ICSME} and \textit{Library Code Generation}~\cite{zhou2023docprompting}, the framework identified semantic concepts as the key information type, overriding the default lexical preference (\textbf{Finding 6}) to utilize \textit{Dense Retrieval}. It further tailored the pipeline by employing query elaboration (\textbf{Finding 2}) and a code-specialized generator (\textbf{Finding 14}) to extract intent for low-clarity refactoring, while retaining the baseline query (\textbf{Finding 4}) to preserve informal user intent in library generation.
    
    \item \textbf{Preserving Structural Precision:} Conversely, for \textit{Test Generation}~\cite{shin2024retrievalaugmentedtestgenerationfar}, the framework detected a high-clarity reliance on explicit structural patterns. It retained the baseline query to prevent semantic drift (\textbf{Findings 1 \& 4}) and recommended \textit{BM25} for exact keyword matching (\textbf{Finding 6}). By omitting re-ranking to protect optimal lexical ordering (\textbf{Finding 10}), the framework's decisions mirrored the original authors' observation that exact matching avoids introducing irrelevant context noise.
\end{itemize}

This confirms our framework possesses robust generalization capabilities, synthesizing empirical findings to autonomously architect theoretically sound pipelines for novel SE tasks.

\begin{table}[htbp]
\centering
\caption{External validation of the adaptive framework across diverse SE tasks, detailing the generated profiles, recommended pipelines, and their rationales mapped to our empirical findings (F denotes Finding).}
\label{tab:external_validation}
\renewcommand{\arraystretch}{1.0}
\resizebox{\textwidth}{!}{
\begin{tabular}{@{} l p{4.5cm} p{4.5cm} p{4.5cm} @{}}
\toprule[1.2pt]
\textbf{Dimension} & \textbf{Move Method Refactoring}~\cite{Bellur2025ICSME} & \textbf{Test Generation}~\cite{shin2024retrievalaugmentedtestgenerationfar} & \textbf{Library Code Gen.}~\cite{zhou2023docprompting} \\
\midrule
\multicolumn{4}{l}{\textit{\textbf{1. Generated Task Profile}}} \\
\midrule
\textbf{Modality} & [Code]-to-[Code] & [Text+Code]-to-[Code] & [Text]-to-[Code] \\
\textbf{Clarity} & Low & High & Medium \\
\textbf{Complexity} & High & Medium & High \\
\textbf{Info Type} & Semantic Concepts & Structural Patterns & Semantic Concepts \\
\textbf{Metric} & Structural Similarity & Functional Correctness & Functional Correctness \\
\midrule
\multicolumn{4}{l}{\textit{\textbf{2. Recommended Pipeline}}} \\
\midrule
\textbf{Query Proc.} & Elaborated & Baseline & Baseline \\
\textbf{Retrieval} & Dense & BM25 & Dense \\
\textbf{Refinement} & None & None & None \\
\textbf{Generator} & Code-Specialized & Frontier Generalist & Frontier Generalist \\
\midrule
\multicolumn{4}{l}{\textit{\textbf{3. Rationale \& Empirical Alignment}}} \\
\midrule
\textbf{Query Logic} & \textbf{F2:} Elaborates ambiguous code smells to extract structural intent. & \textbf{F1, F4:} Retains baseline query for high-clarity tasks to prevent semantic drift. & \textbf{F4:} Avoids rewriting to preserve informal user intent and prevent information loss. \\
\addlinespace
\textbf{Retrieve Logic} & \textbf{F6:} Overrides default lexical search; uses dense embeddings for conceptual bridging. & \textbf{F6:} Relies on exact keyword matching for precise method signatures. & \textbf{F6:} Employs dense retrieval to resolve severe vocabulary mismatches. \\
\addlinespace
\textbf{Other Logic} & \textbf{F10, F14:} Omits re-ranking; selects specialized generator for deep code comprehension. & \textbf{F10, F15:} Preserves lexical ordering (no re-ranking) and limits context size to avoid noise. & \textbf{F10, F14:} Omits refinement; allocates frontier model for complex generative reasoning. \\
\bottomrule[1.2pt]
\end{tabular}}
\end{table}

\subsection{Retrieval Dynamics: Depth and Source Distribution}

\subsubsection{The Impact of Retrieval Depth ($k$)}
\label{sec:discussion_k_value}

\begin{figure}[!ht]
    \centering
    \includegraphics[width=\textwidth]{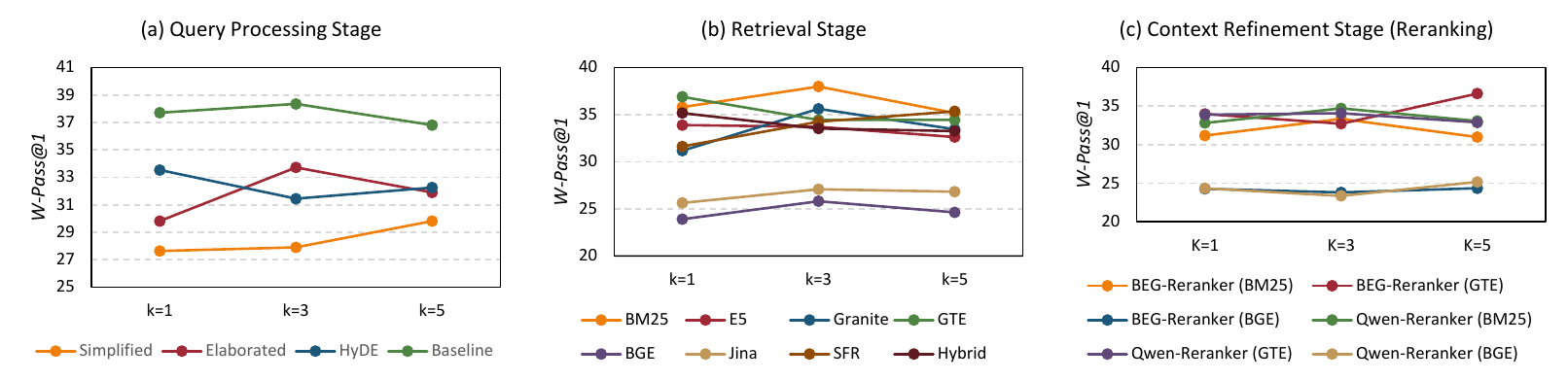}
    \caption{Impact of the number of retrieved documents ($k$) on \textit{W-Pass@1} for different RAG stages on the APPS.}
    \Description{Impact of the number of retrieved documents ($k$) on \textit{W-Pass@1} for different RAG stages on the APPS.}
    \label{fig:k_impact_apps}
\end{figure}

Our analysis of the number of retrieved documents ($k$) reveals a trade-off between signal and noise. As illustrated for the APPS task in \autoref{fig:k_impact_apps}, the relationship between $k$ and performance is not linear. For most components, performance improves significantly from $k=1$ to $k=3$, highlighting the need for a sufficiently rich context. However, performance typically plateaus or declines when moving to $k=5$, as additional documents are more likely to introduce distracting noise. This is especially pronounced for the top-performing retriever, \textit{BM25}.

The optimal $k$ is not universal. For instance, the condensed \textit{Simplified} query uniquely benefits up to $k=5$ to gather sufficient information, while \textit{Re-ranking} renders $k$'s impact unpredictable. This reinforces a key design principle for Code RAG: \textbf{retrieval depth ($k$) must be carefully optimized, not blindly maximized, to balance essential signal against overwhelming noise.}

\subsubsection{Retrieval Source Ablation}
\label{sec:source_ablation}

To validate our heterogeneous corpus design, we analyze the retrieval source distribution at $k=3$. \autoref{tab:source_distribution} demonstrates that no single source universally satisfies the diverse queries of code tasks.

\begin{table}[htbp]
\centering
\caption{Retrieval source distribution at $k=3$. Percentages indicate the proportion of documents retrieved from each corpus source across the evaluated tasks.}
\label{tab:source_distribution}
\resizebox{\linewidth}{!}{
\begin{tabular}{ll ccccc}
\toprule[1.2pt]
\textbf{Task} & \textbf{Dataset} & \textbf{Python API} & \textbf{Stack Overflow} & \textbf{CodeSearchNet} & \textbf{Code-Contests} & \textbf{LeetCode} \\
\midrule
Code Gen. & APPS & 1.41\% & 2.81\% & 0.44\% & \textbf{83.83\%} & 11.50\% \\
\midrule
\multirow{2}{*}{Code Sum.} & CodeXGLUE & 31.85\% & 24.00\% & \textbf{39.00\%} & 4.77\% & 0.38\% \\
& CodeXGLUE-OBF & 14.47\% & 29.15\% & \textbf{43.41\%} & 10.10\% & 2.88\% \\
\midrule
Code Repair & DebugBench & 1.34\% & 2.69\% & 1.31\% & 14.03\% & \textbf{80.63\%} \\
\bottomrule[1.2pt]
\end{tabular}}
\end{table}

Task-specific dominance is quantitatively evident. Code generation retrieves 83.83\% of its context from Code-Contests, suggesting models may heavily rely on competitive programming boilerplate to synthesize complex solutions. Conversely, code repair draws 80.63\% from LeetCode, indicating that logic debugging inherently demands precise algorithmic ground truth. Meanwhile, code summarization exhibits a balanced triangulation: CodeSearchNet provides 39.00\% of structural context, while the Python API and Stack Overflow contribute 31.85\% and 24.00\% respectively to supply functional definitions.

Code obfuscation induces a critical retrieval shift. When explicit identifiers vanish, Python API retrievals drop substantially from 31.85\% to 14.47\%, forcing the system to compensate by extracting syntactic patterns from CodeSearchNet and community discussions from Stack Overflow, which correspondingly increase. These dynamic adaptations confirm that a heterogeneous corpus can effectively handle varying query clarities, proving its construction is a functional necessity rather than an arbitrary design choice.

\subsection{Robustness of Findings: A Cross-Temporal Validation}
\label{sec:temporal_validation}

To determine whether pre-training memorization on legacy benchmarks invalidates our empirical findings, we conducted a cross-temporal trend analysis. We compared the legacy APPS dataset against a zero-contamination validation set from LiveCodeBench (LCB)~\cite{Jain2025lcb}, comprising 100 problems published post-\textbf{January 2025} (with a balanced 1:1:1 difficulty ratio) to strictly bypass the knowledge cutoff of our evaluated models.

\begin{figure}[htbp]
    \centering
    \includegraphics[width=\textwidth]{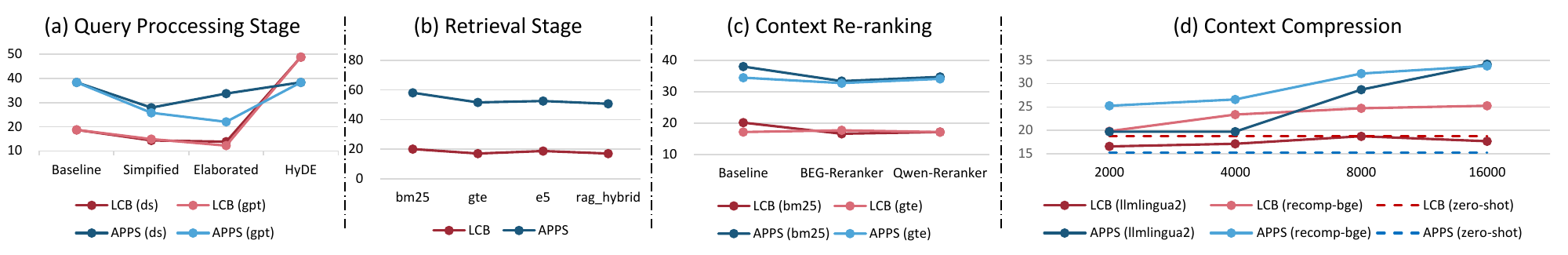}
    \caption{Cross-temporal validation of core RAG findings. The relative performance trends across (a) Query Processing, (b) Retrieval, (c) Context Re-ranking, and (d) Context Compression remain highly consistent.}
    \Description{Cross-temporal validation of core RAG findings. The relative performance trends across (a) Query Processing, (b) Retrieval, (c) Context Re-ranking, and (d) Context Compression remain highly consistent.}
    \label{fig:lcb_vs_apps}
\end{figure}

Crucially, while absolute generation scores naturally degrade on the uncontaminated LCB dataset, the relative ranking of RAG configurations remains remarkably invariant (\autoref{fig:lcb_vs_apps}). \textit{BM25} strictly dominates both dense and hybrid retrievers across both temporal splits (\textbf{Findings 6 \& 9}), proving that the necessity for precise lexical matching in code generation is an intrinsic task property, not a byproduct of retrieving memorized snippets. Heavy query transformations (e.g., \textit{Elaborated}, \textit{HyDE}) consistently underperform \textit{Baseline} and \textit{Simplified} inputs (\textbf{Findings 1 \& 4}), confirming that semantic drift in over-engineered queries penalizes novel problem-solving just as it does on legacy data. Context refinement introduces universal regressions; models like \textit{BGE-Reranker} strictly degrade baseline retrieval accuracy (\textbf{Finding 10}), demonstrating their persistent misalignment with underlying algorithmic logic regardless of data recency. Finally, context compression strategies display a converging upward trajectory with expanding token budgets (\textbf{Finding 12}), corroborating their stable utility as noise filters.

Ultimately, this temporal validation yields a definitive conclusion: \textbf{\textit{data leakage on legacy benchmarks inflates absolute generation scores but does not confound the underlying RAG dynamics}}. The comparative advantages of specific pipeline components are intrinsically tied to the algorithmic nature of the tasks, verifying that our empirical insights generalize robustly to contemporary, unseen SE challenges.

\subsection{Cost and Feasibility Analysis}
\label{sec:cost_analysis}

To assess real-world feasibility, we profile the computational and token overheads across RAG components (\autoref{tab:cost_analysis}) using our standardized infrastructure (\autoref{sec:experimental_environment}).

\textbf{Computational Latency.} Resource demands vary drastically. \textit{BM25} is computationally negligible ($<$10ms), whereas deep-learning bottlenecks are severe. \textit{Re-ranking} (\textit{BGE-Reranker}) consumes 815ms alongside massive VRAM ($\sim$12.8GB), orders of magnitude slower than retrieval. This quantitatively reinforces \textbf{Finding 10}, confirming that indiscriminate re-ranking renders pipelines engineered-infeasible for latency-sensitive SE workflows.

\textbf{Token Economics.} Cost is driven by prompt inflation. \textit{Elaborated} queries incur a heavy 987-token overhead, roughly equivalent to retrieving 2-3 extra documents, explicitly justifying our restriction to low-clarity tasks (\textbf{Finding 2}). Furthermore, tripling the input volume with $k=3$ (e.g., 1,760 $\rightarrow$ 5,781 in CodeXGLUE) leaves output length stable. Thus, Code RAG costs scale predictably with context ingestion, avoiding erratic generation bloat.

\begin{table}[htbp]
\centering
\caption{Cost and feasibility profiling. \textbf{Left:} Computational overhead per query (Ret.: Retrival, BGE-R: BGE-Reranker). \textbf{Right:} Token consumption dynamics (Simp.: Simplified, Elab.: Elaborated, CXG: CodeXGLUE, DB: DebugBench, ``-'': non-applicability).}
\label{tab:cost_analysis}
\renewcommand{\arraystretch}{1.1}
\setlength{\tabcolsep}{3pt}
\resizebox{\linewidth}{!}{
\begin{tabular}{@{}llcc@{}}
\toprule[1.2pt]
\multicolumn{4}{c}{\textit{\textbf{Panel A: Computational Overhead}}} \\
\midrule
\textbf{Stage} & \textbf{Method} & \textbf{Device (V/RAM)} & \textbf{Latency} \\
\midrule
Sparse Ret. & \textit{BM25} & CPU ($\sim$1.5G) & $<$10ms \\
Dense Ret. & \textit{E5 (118M)} & GPU ($\sim$3.5G) & 205ms \\
Re-ranking & \textit{BGE-R} & GPU ($\sim$12.8G) & 815ms \\
\bottomrule[1.2pt]
\end{tabular}

\hspace{0.5cm}
\begin{tabular}{@{}llcc@{}}
\toprule[1.2pt]
\multicolumn{4}{c}{\textit{\textbf{Panel B: Average Token Consumption}}} \\
\midrule
\textbf{Component} & \textbf{Setting} & \textbf{Input Tokens} & \textbf{Output Tokens} \\
\midrule
Query Procession & \textit{Simp.} / \textit{Elab.} & +85 / +987 & - \\
Generator ($k=0 \rightarrow 3$) & APPS & 740 $\rightarrow$ 2,926 & 266 $\rightarrow$ 217 \\
Generator ($k=0 \rightarrow 3$) & CXG & 1,760 $\rightarrow$ 5,781 & 111 $\rightarrow$ 109 \\
Generator ($k=0 \rightarrow 3$) & DB & 845 $\rightarrow$ 2,518 & 234 $\rightarrow$ 187 \\
\bottomrule[1.2pt]
\end{tabular}
}
\end{table}

\subsection{Threats to Validity}
\label{sec:discussion_threats}

\noindent\textbf{External Validity:}
Our findings are based on the Python language and three specific tasks, which may limit generalizability. We partially mitigate the risk of dataset-specific conclusions by evaluating diverse query transformations in \autoref{sec:rq1}. These transformations simulate the diverse input patterns one might encounter across different datasets for the same task. Furthermore, the rapidly evolving LLM landscape means our model selection is a snapshot in time; future models may exhibit different sensitivities to the RAG components we studied.

\noindent\textbf{Construct Validity:}
Our metrics, while standard, have limitations. \textit{W-Pass@1} evaluates only functional correctness, ignoring readability and efficiency. Additionally, lower \textit{CodeBLEU} scores in code repair don't necessarily indicate RAG failure; rather, RAG often prioritizes functionally correct external solutions over structurally minimal edits.

\noindent\textbf{Internal Validity:}
We mitigated errors by leveraging well-established libraries and pre-trained models. The primary source of potential internal risk lies in the custom code written to connect these disparate components into a cohesive experimental pipeline. While this ``glue code'' was carefully tested, the possibility of subtle, undetected bugs cannot be entirely eliminated.

\section{Related Work}
\label{sec:related_work}

\noindent\textbf{Retrieval-Augmented Generation for SE.} 
The application of RAG to SE is a burgeoning field, with research demonstrating its value across various tasks. Early works established that retrieving relevant code snippets significantly improves code completion and summarization~\cite{parvez2021retrieval, lu-etal-2022-reacc}. Subsequent research has explored more sophisticated retrieval sources beyond simple code snippets. For instance, DocPrompting showcased the effectiveness of retrieving from API documentation to handle libraries unseen during training~\cite{zhou2023docprompting}, while others have proposed using structured knowledge, such as representing a codebase as a knowledge graph, to enhance retrieval for repository-level tasks~\cite{Athale_2025}. More recent work has even begun to explore RAG for specialized domains like test generation~\cite{shin2024retrievalaugmentedtestgenerationfar}. As categorized in recent surveys~\cite{gao2024retrievalaugmented}, these studies typically propose and validate a single, novel RAG architecture or component. While pioneering new methods, they do not provide a comparative analysis of the interchangeable components (e.g., different retrievers, re-rankers) that constitute the pipeline. Our work differs fundamentally by providing the first systematic, component-wise empirical study of the RAG design space for code.

\noindent\textbf{Empirical Studies and Benchmarking of LLMs for Code.}
Our research also builds upon the strong tradition of empirical evaluation in SE and AI. The advancement of LLMs for code has been driven by standardized benchmarks that enable rigorous comparison, such as APPS for code generation~\cite{hendrycks2021measuring}, CodeXGLUE for a variety of code intelligence tasks~\cite{lu2021codexglue}, and HumanEval for evaluating foundational model capabilities~\cite{chen2021evaluating}. Within this context, several studies have conducted ablation or comparative analyses on specific aspects of the code generation process, such as the impact of different prompting techniques~\cite{Wei2022ChainOfThought} or the effect of model fine-tuning versus prompting~\cite{Wang2022Nomore}. However, these empirical studies have largely focused on the LLM itself, treating it as a monolithic component. To the best of our knowledge, no prior work has applied this rigorous, empirical methodology to systematically dissect and evaluate the multi-stage RAG pipeline specifically for code-related tasks. Our study fills this critical gap by providing a foundational empirical roadmap that connects task features to optimal RAG pipeline configurations.

\section{Conclusion}
\label{sec:conclusion}

Our large-scale, component-wise study of RAG for software engineering leads to a clear directive: practitioners should \textbf{prioritize the retriever} and embrace \textbf{architectural parsimony}. We found that pipeline performance is dominated by the retrieval stage, where classic lexical models like \textit{BM25} often outperform larger dense counterparts. Advanced components like query transformers and re-rankers, in contrast, provide only situational benefits and can even be detrimental. Furthermore, our results underscore the highly task-dependent nature of the entire pipeline---from retriever to generator---challenging the efficacy of static, one-size-fits-all systems. This work therefore strongly advocates for a paradigm shift towards the \textbf{adaptive, task-aware frameworks} we prototype. Ultimately, our study provides a foundational empirical roadmap for engineering RAG systems for code in a more principled, data-driven manner.

\section*{Data Availability} 
The artifact of this paper, containing all experimental data, source code, and documentation necessary to reproduce our component-wise empirical study of RAG systems for software engineering tasks, can be accessed via \url{https://github.com/security-pride/RAG-Empirical-SE}.

\bibliographystyle{ACM-Reference-Format}
\bibliography{main}

\end{document}